# Prediction of PLX-4720 Sensitivity in Cancer Cell Lines through Multi-Omics Integration and Attention-Based Fusion Modeling

La Ode Aman[1*], Arfan Arfan[2], Aiyi Asnaw[3], Purnawan Pontana Putra[4], Hamsidar Hasan[1], Dizky Ramadani Putri Papeo[1]

[1] Department of Pharmacy, Faculty of Sports and Health, Universitas Negeri Gorontalo, Jl. Jend. Sudirman No.6, Dulalowo Timur., Kec. Kota Tengah, Gorontalo, 96128, Indonesia
[2] Department of Pharmaceutical Analysis and Medicinal Chemistry, Universitas Halu Oleo, Jl. H.E.A. Mokodompit, Kendari, 93231, Southeast Sulawesi, Indonesia
[3] Department of Pharmacochemistry, Faculty of Pharmacy, Universitas Bhakti Kencana, Jl. Soekarno-Hatta No. 754, Bandung, 40617, Indonesia
[4] Department of Pharmaceutical Chemistry, Faculty of Pharmacy, Universitas Andalas, Padang, 25163, Indonesia

*Corresponding author: laode_aman@ung.ac.id

## ABSTRACT

Predicting the sensitivity of cancer cell lines to PLX-4720, a preclinical BRAF inhibitor, requires models capable of capturing the multilayered regulation of oncogenic signaling. Single-omics predictors are often insufficient because drug response is shaped by interactions among genomic alterations, epigenetic regulation, transcriptional activity, protein signaling, metabolic state, and network-level context. In this study we develop an attention-based multi-omics integration framework using genomic, epigenomic, transcriptomic, proteomic, metabolomic, and protein interaction data from the GDSC1 panel. Each modality is encoded into a latent representation using feed-forward neural networks or graph convolutional networks, and fused through an attention mechanism that assigns modality-specific importance weights. A regression model is then used to predict PLX-4720 response. Across single- and multi-omics configurations, the best performance is achieved by integrating genomics and transcriptomics, which yields validation R2 values above 0.92. This reflects the complementary roles of mutational status and downstream transcriptional activation in shaping sensitivity to BRAF inhibition. Epigenomics is the strongest single-omics predictor, while metabolomics and PPI data contribute additional context when combined with other modalities. Integration of three to five omics layers improves stability but does not surpass the accuracy of the best two-modality combinations, likely due to information redundancy and sample-size imbalance. These findings highlight the importance of modality selection rather than maximal data depth. The proposed framework provides an efficient and biologically grounded strategy for drug response prediction and supports the development of precision pharmacogenomics.

# INTRODUCTION

Molecularly targeted cancer therapies have advanced significantly following the discovery of BRAF inhibitors capable of suppressing constitutive MAPK pathway activation driven by the BRAF V600E mutation. PLX-4720 is a preclinical compound that selectively inhibits BRAF and serves as a precursor to the clinical drug Vemurafenib, which has been approved by the FDA for melanoma harboring the BRAF V600E mutation (Bollag et al., 2010; Flaherty et al., 2010). The BRAF V600E mutation triggers abnormal activation of the MAPK/ERK pathway, making tumor cells dependent on BRAF activity for proliferation—a condition known as oncogene addiction (Wellbrock & Arozarena, 2016). Consequently, drug response to BRAF inhibitors is heavily influenced by upstream molecular mechanisms and the complex regulation of signaling pathways.

Although several studies have demonstrated the effectiveness of BRAF inhibitors in specific subsets of tumor cell lines, drug sensitivity prediction based on a single layer of molecular data remains limited. For instance, genomic data alone cannot fully explain the heterogeneity of responses across cell lines, even when identical driver mutations are present (Geeleher, Cox, & Huang, 2014). Epigenetic factors, gene expression, protein signaling, and metabolic context simultaneously contribute to modulating drug responses, rendering the determinants of BRAF inhibitor sensitivity multilevel and nonlinear.

Over the past decade, multi-omics integration strategies have emerged as a powerful approach for understanding the determinants of drug sensitivity, as they capture biological variations that are invisible to single-modality analyses (Hasin, Seldin, & Lusis, 2017). Integration of genomics, epigenomics, transcriptomics, proteomics, metabolomics, and protein interaction networks has been shown to enhance our understanding of resistance mechanisms and responses to MAPK inhibitors across various cancer types (Zhang et al., 2021). However, traditional integration methods often rely on concatenation, matrix factorization, or cross-modality correlations, which are limited in their ability to accommodate feature heterogeneity and fail to capture nonlinear relationships across different layers of biological regulation.

Recent advances in artificial intelligence have enabled the development of attention-based fusion approaches, in which each omics modality is projected into a shared representation space and weighted dynamically based on its informational contribution (Peng et al., 2022). The attention mechanism allows the model to automatically learn which modalities are most relevant to the pharmacological response of each sample, producing more stable and meaningful latent representations than standard concatenation (Chaudhary, Poirion, Lu, & Garmire, 2018). This approach is particularly valuable in pharmacogenomics, where drug sensitivity represents a complex phenotype involving interplay between multiple layers of molecular regulation.

In the context of PLX-4720, integrated baseline multi-omics data can explain BRAF inhibitor sensitivity more accurately than any single modality, as drug response is influenced by upstream factors such as BRAF/RAS mutations, epigenetic regulation, MAPK protein expression, metabolic activity, and the structure of protein signaling networks. Therefore, attention-based multi-omics integration represents a highly relevant approach for modeling PLX-4720 sensitivity in cancer cell lines, particularly to strengthen preclinical pharmacogenomic precision and support more effective selection of targeted therapies.

# METHODS

## Study Design and Data Sources

This study is an *in silico* investigation employing a retrospective observational design based on pharmacogenomic data from cancer cell lines. Drug sensitivity data and multi-omics molecular profiles were obtained from the **Genomics of Drug Sensitivity in Cancer (GDSC)** database, which provides drug response information and multiple layers of molecular data for hundreds of cancer cell lines across diverse tumor types (Iorio et al., 2016; Yang et al., 2013).

The analysis focused on **PLX-4720 (GDSC1:1371)**, a preclinical BRAF inhibitor. Samples consisted of cancer cell lines with available PLX-4720 response data as well as corresponding genomic, epigenomic, transcriptomic, proteomic, metabolomic, and protein–protein interaction (PPI) profiles. Only cell lines with complete data or, at minimum, valid features–label pairs for PLX-4720 were included in the analysis.

## Multi-Omics Data

### *Drug Response Data*

PLX-4720 response data were extracted in the form IC50 values, following the GDSC data format. These values were used as continuous targets for regression-based drug sensitivity modeling, reflecting the degree of cell line sensitivity to PLX-4720 (Iorio et al., 2016).

### *Genomic Data*

Genomic data comprised profiles of single nucleotide variants (SNVs) and small insertions/deletions (indels) in key cancer-related genes. Genomic features were represented as binary matrices (mutation present/absent) per gene per cell line, consistent with commonly used approaches in genomics-based drug response prediction studies (Geeleher, Cox, & Huang, 2014; Liu et al., 2021).

### *Epigenomic Data*

Epigenomic data were derived from DNA methylation profiles at CpG sites or gene promoter regions, reflecting transcriptional regulatory states. β-values or M-values were processed into continuous features per cell line, following established practices for methylation integration in multi-omics cancer studies (Hoadley et al., 2018).

***Transcriptomic Data***

Transcriptomic data consisted of gene expression measurements (e.g., TPM/FPKM or normalized microarray intensities) and were represented as gene-by-cell-line matrices. Only genes with sufficient variability and non-constant expression across cell lines were retained, in accordance with standard practices to reduce dimensionality and noise (Rappoport & Shamir, 2018).

***Proteomic Data***

Proteomic data included protein or phosphoprotein expression profiles measured across cancer cell line panels. Intensity values or relative expression levels were normalized and used as continuous features capturing the functional state of major signaling pathways, which are often more proximal to drug response phenotypes than gene expression levels (Akbani et al., 2014).

***Metabolomic Data***

Metabolomic data comprised relative concentrations of key cellular metabolites, reflecting metabolic states and metabolic rewiring associated with MAPK pathway dependency and responses to BRAF inhibition (DeBerardinis & Chandel, 2020). These data were compiled as metabolite-by-cell-line matrices.

***Protein–Protein Interaction (PPI) Network Data***

PPI data were constructed from the STRING database or other widely used protein network resources, with edges selected based on high-confidence interaction scores (Szklarczyk et al., 2021). Nodes represent proteins, while edge weights reflect interaction strength or functional evidence. Node features were initialized from protein expression or abundance data and subsequently processed using a Graph Convolutional Network (GCN) to generate node-level PPI embeddings, which were then pooled per cell line to produce sample-specific PPI representations.

## Data Preprocessing

All feature matrices for each modality were processed using the following standard steps:

- **Handling missing values:** Missing values were imputed using the per-feature median or set to zero after normalization when distributional properties permitted, following robust preprocessing practices (Hasin, Seldin, & Lusis, 2017).

- **Removal of constant and duplicate features:** Features with near-zero variance ($< 1e{-}8$) and duplicate features were removed to reduce redundancy and prevent training instability.

- **Scale normalization:** Continuous features were normalized using RobustScaler or column-wise z-score normalization to mitigate the influence of outliers and harmonize feature scales, in line with preprocessing recommendations for large-scale omics data (Bersanelli et al., 2016).

- **Sample ID harmonization:** All cell line identifiers were standardized (upper case, trimming) to ensure consistency across modalities and drug response files, enabling accurate cross-modality mapping.

## Multi-Omics Embedding Construction

### *Modality-Specific Encoders*

Each omics modality was processed using a feed-forward neural network encoder (the encoder component of an autoencoder) that projected high-dimensional feature vectors into lower-dimensional latent spaces (e.g., 128 dimensions). Encoder architectures consisted of sequential linear layers with Layer Normalization, ReLU/GELU activation functions, and dropout, following established practices for multi-omics embedding construction (Chaudhary, Poirion, Lu, & Garmire, 2018; Zhang et al., 2021). Encoders were trained to produce stable representations free of NaN/Inf values using safe forward procedures.

The output of each encoder was stored as an omics-specific embedding matrix of shape [n_samples × latent_dim] and used as input for the integration stage.

### *PPI Network Embedding*

For PPI data, node features (based on protein expression) were projected using a two-layer Graph Convolutional Network (GCN) to generate protein-level embeddings. The GCN aggregates local neighborhood information from the interaction network, enabling representations that capture pathway context and protein modules (Kipf & Welling, 2017).

Node embeddings were subsequently pooled per cell line using top-k or expression-weighted averaging strategies, yielding a single PPI vector per sample that reflects the functional state of the protein interaction network.

## Attention-Based Multi-Omics Integration

### *Attention-Based Fusion Scheme*

Cross-modality integration was performed using an attention-based fusion model. Embeddings from each modality $z^{(m)}$ were first projected into a shared latent space of dimension ATTN_DIM via linear layers. An attention mechanism then computed modality-specific importance scores for each sample and generated normalized softmax weights. The integrated representation was obtained as a weighted sum of modality embeddings:

$$h = \sum_{m} \alpha_{m} z^{(m)}$$

This approach enables the model to adaptively learn which modalities are most relevant for predicting PLX-4720 sensitivity in each cancer cell line, as demonstrated in prior multi-omics integration studies for cancer subtype and outcome prediction (Chaudhary et al., 2018; Peng et al., 2022).

***Prediction Head (Regression Head)***

The integrated representation h\mathbf{h}h was fed into a regression head implemented as a multilayer perceptron (MLP) with one or two hidden layers and non-linear activation functions, followed by a single linear output layer to produce continuous PLX-4720 response predictions (e.g., AUC or ln(IC50)). Regularization was applied using dropout and weight decay to prevent overfitting.

## 6. Model Training and Validation Scheme

Models were trained using the Adam optimizer with an initial learning rate of approximately $1 \times 10^{-3}$, a batch size of 128, and Mean Squared Error (MSE) as the primary loss function between predicted and observed values. The dataset was split into training, validation, and test sets with an approximate ratio of 60–20–20, preserving the distribution of drug response values.

During training, early stopping was applied based on validation performance with a patience of several epochs to prevent overfitting. The best-performing model parameters (based on lowest validation MSE or best generalization metrics) were saved and used for final evaluation on the test set. This strategy is consistent with common practices in deep learning–based multi-omics modeling and drug response prediction (Sharifi-Noghabi et al., 2020; Zhang et al., 2021).

## 7. Performance Evaluation and Modality Combination Experiments

Model performance was evaluated using regression metrics, primarily test loss (MSE on the test set) and indicators of generalization stability (differences or ratios among training, validation, and test losses). In addition, a series of experiments were conducted involving:

- **Single-modality models:** Models trained using embeddings from a single modality (e.g., genomics only, proteomics only) to assess the informative contribution of each molecular layer.

- **Incremental multi-omics combinations:** Models trained using various combinations of two, three, four, and five modalities, including the full configuration integrating genomics, epigenomics, transcriptomics, proteomics, metabolomics, and PPI data.

- **Performance comparison:** Each combination was evaluated based on test loss and generalization stability. The combination with the lowest test loss and highest stability was interpreted as the most informative multi-omics configuration for predicting PLX-4720 sensitivity.

This analysis not only assessed predictive performance but also provided insights into which layers of molecular regulation contribute most strongly to PLX-4720 response in cancer cell lines, in alignment with the goals of precision pharmacogenomics and multi-omics integration (Hasin et al., 2017; Sharifi-Noghabi et al., 2020).

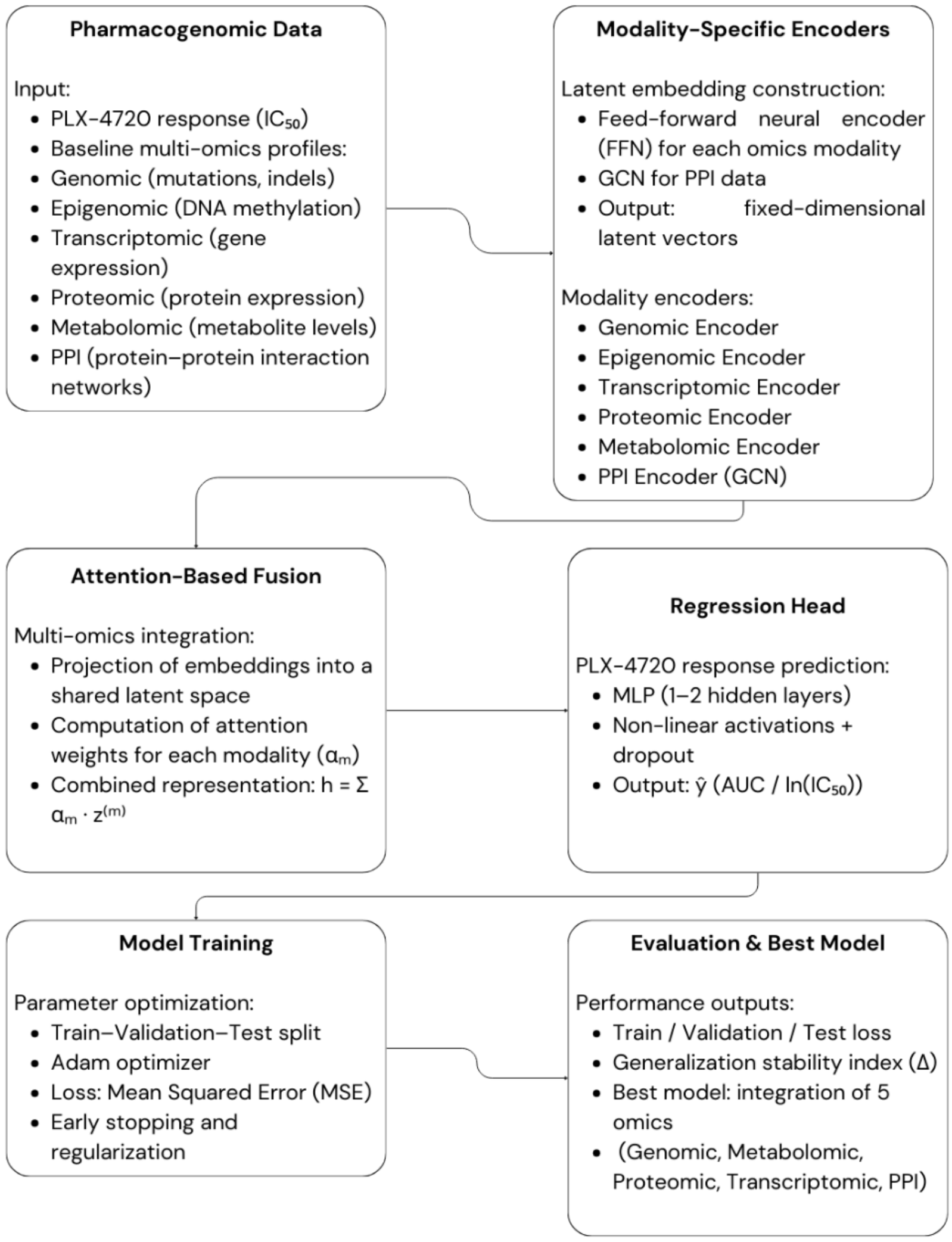

**Figure 1.** Attention-based multi-omics integration workflow for predicting PLX-4720 sensitivity in cancer cell lines. Baseline multi-omics data are mapped to modality-specific latent embeddings, fused using an attention-based mechanism, and projected through a regression head to predict drug response.

## RESULT

Table 1. Performance of single- and multi-omics models for predicting PLX-4720 sensitivity.

| drug | modalities | n_feat_samples | n_labeled | MSE | RMSE | MAE | $R^2$ |
|---|---|---|---|---|---|---|---|
| PLX 4720 (GDSC1:1036) | Metabolomics | 632 | 30 | 3.31 | 1.82 | 1.59 | -0.13 |
| PLX 4720 (GDSC1:1036) | Proteomics | 620 | 30 | 2.90 | 1.70 | 1.45 | 0.02 |
| PLX 4720 (GDSC1:1036) | Genomics | 949 | 45 | 2.32 | 1.52 | 1.31 | 0.24 |
| PLX 4720 (GDSC1:1036) | Epigenomics | 559 | 28 | 1.50 | 1.23 | 0.92 | 0.46 |
| PLX 4720 (GDSC1:1036) | Transcriptomics | 708 | 34 | 3.03 | 1.74 | 1.52 | -0.11 |
| PLX 4720 (GDSC1:1036) | PPI | 708 | 34 | 2.53 | 1.59 | 1.32 | 0.08 |
| PLX 4720 (GDSC1:1036) | Metabolomics + Proteomics | 615 | 28 | 2.24 | 1.50 | 1.21 | 0.22 |
| PLX 4720 (GDSC1:1036) | Genomics + Metabolomics | 630 | 30 | 2.40 | 1.55 | 1.34 | 0.18 |
| PLX 4720 (GDSC1:1036) | Epigenomics + Metabolomics | 554 | 27 | 1.14 | 1.07 | 0.83 | 0.60 |
| PLX 4720 (GDSC1:1036) | Metabolomics + Transcriptomics | 622 | 29 | 0.26 | 0.51 | 0.31 | 0.91 |
| PLX 4720 (GDSC1:1036) | Metabolomics + PPI | 622 | 29 | 0.37 | 0.61 | 0.45 | 0.87 |
| PLX 4720 (GDSC1:1036) | Genomics + Proteomics | 618 | 30 | 1.51 | 1.23 | 0.85 | 0.49 |
| PLX 4720 (GDSC1:1036) | Epigenomics + Proteomics | 544 | 28 | 1.00 | 1.00 | 0.67 | 0.64 |
| PLX 4720 (GDSC1:1036) | Proteomics + Transcriptomics | 613 | 29 | 4.09 | 2.02 | 1.74 | -0.44 |
| PLX 4720 (GDSC1:1036) | PPI + Proteomics | 613 | 29 | 4.07 | 2.02 | 1.74 | -0.43 |
| PLX 4720 (GDSC1:1036) | Epigenomics + Genomics | 558 | 28 | 1.95 | 1.40 | 1.10 | 0.30 |
| PLX 4720 (GDSC1:1036) | Genomics + Transcriptomics | 707 | 34 | 0.19 | 0.44 | 0.28 | 0.93 |
| PLX 4720 (GDSC1:1036) | Genomics + PPI | 707 | 34 | 1.19 | 1.09 | 0.87 | 0.57 |
| PLX 4720 (GDSC1:1036) | Epigenomics + Transcriptomics | 553 | 28 | 1.01 | 1.01 | 0.75 | 0.64 |
| PLX 4720 (GDSC1:1036) | Epigenomics + PPI | 553 | 28 | 1.05 | 1.02 | 0.67 | 0.63 |
| PLX 4720 (GDSC1:1036) | PPI + Transcriptomics | 708 | 34 | 2.31 | 1.52 | 1.17 | 0.16 |
| PLX 4720 (GDSC1:1036) | Genomics + Metabolomics + Proteomics | 613 | 28 | 1.55 | 1.24 | 0.96 | 0.46 |
| PLX 4720 (GDSC1:1036) | Epigenomics + Metabolomics + Proteomics | 541 | 27 | 2.30 | 1.52 | 1.23 | 0.18 |
| PLX 4720 (GDSC1:1036) | Metabolomics + Proteomics + Transcriptomics | 609 | 28 | 1.07 | 1.03 | 0.58 | 0.63 |
| PLX 4720 (GDSC1:1036) | Metabolomics + PPI + Proteomics | 609 | 28 | 1.36 | 1.16 | 0.72 | 0.53 |
| PLX 4720 (GDSC1:1036) | Epigenomics + Genomics + Metabolomics | 553 | 27 | 0.74 | 0.86 | 0.56 | 0.74 |
| PLX 4720 (GDSC1:1036) | Genomics + Metabolomics + Transcriptomics | 621 | 29 | 1.20 | 1.09 | 0.90 | 0.59 |
| PLX 4720 (GDSC1:1036) | Genomics + Metabolomics + PPI | 621 | 29 | 0.93 | 0.97 | 0.76 | 0.68 |
| PLX 4720 (GDSC1:1036) | Epigenomics + Metabolomics + Transcriptomics | 548 | 27 | 0.68 | 0.82 | 0.45 | 0.76 |
| PLX 4720 (GDSC1:1036) | Epigenomics + Metabolomics + PPI | 548 | 27 | 0.93 | 0.97 | 0.69 | 0.67 |
| PLX 4720 (GDSC1:1036) | Metabolomics + PPI + Transcriptomics | 622 | 29 | 1.88 | 1.37 | 1.13 | 0.36 |
| PLX 4720 (GDSC1:1036) | Epigenomics + Genomics + Proteomics | 543 | 28 | 1.17 | 1.08 | 0.68 | 0.58 |
| PLX 4720 (GDSC1:1036) | Genomics + Proteomics + Transcriptomics | 612 | 29 | 3.69 | 1.92 | 1.66 | -0.29 |
| PLX 4720 (GDSC1:1036) | Genomics + PPI + Proteomics | 612 | 29 | 4.39 | 2.09 | 1.81 | -0.54 |
| PLX 4720 (GDSC1:1036) | Epigenomics + Proteomics + Transcriptomics | 539 | 28 | 0.74 | 0.86 | 0.57 | 0.74 |
| PLX 4720 (GDSC1:1036) | Epigenomics + PPI + Proteomics | 539 | 28 | 2.46 | 1.57 | 1.34 | 0.12 |
| PLX 4720 (GDSC1:1036) | PPI + Proteomics + Transcriptomics | 613 | 29 | 3.91 | 1.98 | 1.71 | -0.37 |
| PLX 4720 (GDSC1:1036) | Epigenomics + Genomics + Transcriptomics | 553 | 28 | 1.06 | 1.03 | 0.73 | 0.62 |
| PLX 4720 (GDSC1:1036) | Epigenomics + Genomics + PPI | 553 | 28 | 0.91 | 0.95 | 0.78 | 0.68 |
| PLX 4720 (GDSC1:1036) | Genomics + PPI + Transcriptomics | 707 | 34 | 1.22 | 1.10 | 0.74 | 0.56 |
| PLX 4720 (GDSC1:1036) | Epigenomics + PPI + Transcriptomics | 553 | 28 | 1.08 | 1.04 | 0.71 | 0.61 |
| PLX 4720 (GDSC1:1036) | Epigenomics + Genomics + Metabolomics + Proteomics | 540 | 27 | 0.88 | 0.94 | 0.62 | 0.69 |
| PLX 4720 (GDSC1:1036) | Genomics + Metabolomics + Proteomics + Transcriptomics | 608 | 28 | 0.86 | 0.93 | 0.49 | 0.70 |
| PLX 4720 (GDSC1:1036) | Genomics + Metabolomics + PPI + Proteomics | 608 | 28 | 2.25 | 1.50 | 1.25 | 0.22 |
| PLX 4720 (GDSC1:1036) | Epigenomics + Metabolomics + Proteomics + Transcriptomics | 536 | 27 | 1.07 | 1.04 | 0.71 | 0.62 |
| PLX 4720 (GDSC1:1036) | Epigenomics + Metabolomics + PPI + Proteomics | 536 | 27 | 0.96 | 0.98 | 0.62 | 0.66 |
| PLX 4720 (GDSC1:1036) | Metabolomics + PPI + Proteomics + Transcriptomics | 609 | 28 | 1.17 | 1.08 | 0.73 | 0.59 |
| PLX 4720 (GDSC1:1036) | Epigenomics + Genomics + Metabolomics + Transcriptomics | 548 | 27 | 1.46 | 1.21 | 0.93 | 0.48 |
| PLX 4720 (GDSC1:1036) | Epigenomics + Genomics + Metabolomics + PPI | 548 | 27 | 1.17 | 1.08 | 0.81 | 0.58 |
| PLX 4720 (GDSC1:1036) | Genomics + Metabolomics + PPI + Transcriptomics | 621 | 29 | 1.71 | 1.31 | 1.02 | 0.42 |
| PLX 4720 (GDSC1:1036) | Epigenomics + Metabolomics + PPI + Transcriptomics | 548 | 27 | 1.18 | 1.09 | 0.83 | 0.58 |
| PLX 4720 (GDSC1:1036) | Epigenomics + Genomics + Proteomics + Transcriptomics | 539 | 28 | 1.29 | 1.14 | 0.87 | 0.54 |
| PLX 4720 (GDSC1:1036) | Epigenomics + Genomics + PPI + Proteomics | 539 | 28 | 2.17 | 1.47 | 1.21 | 0.22 |
| PLX 4720 (GDSC1:1036) | Genomics + PPI + Proteomics + Transcriptomics | 612 | 29 | 3.95 | 1.99 | 1.72 | -0.39 |
| PLX 4720 (GDSC1:1036) | Epigenomics + PPI + Proteomics + Transcriptomics | 539 | 28 | 1.32 | 1.15 | 0.78 | 0.53 |
| PLX 4720 (GDSC1:1036) | Epigenomics + Genomics + PPI + Transcriptomics | 553 | 28 | 0.76 | 0.87 | 0.65 | 0.73 |
| PLX 4720 (GDSC1:1036) | Epigenomics + Genomics + Metabolomics + Proteomics + Transcriptomics | 536 | 27 | 0.69 | 0.83 | 0.52 | 0.75 |
| PLX 4720 (GDSC1:1036) | Epigenomics + Genomics + Metabolomics + PPI + Proteomics | 536 | 27 | 2.88 | 1.70 | 1.42 | -0.02 |
| PLX 4720 (GDSC1:1036) | Genomics + Metabolomics + PPI + Proteomics + Transcriptomics | 608 | 28 | 0.77 | 0.88 | 0.52 | 0.73 |
| PLX 4720 (GDSC1:1036) | Epigenomics + Metabolomics + PPI + Proteomics + Transcriptomics | 536 | 27 | 1.42 | 1.19 | 0.88 | 0.50 |
| PLX 4720 (GDSC1:1036) | Epigenomics + Genomics + Metabolomics + PPI + Transcriptomics | 548 | 27 | 0.69 | 0.83 | 0.51 | 0.76 |
| PLX 4720 (GDSC1:1036) | Epigenomics + Genomics + PPI + Proteomics + Transcriptomics | 539 | 28 | 0.40 | 0.63 | 0.46 | 0.86 |
| PLX 4720 (GDSC1:1036) | Epigenomics + Genomics + Metabolomics + PPI + Proteomics + Transcriptomics | 536 | 27 | 1.07 | 1.03 | 0.75 | 0.62 |

Table 1 summarizes the performance of PLX-4720 sensitivity prediction models across various multi-omics data integration configurations in the GDSC1 cancer cell line panel. Each row indicates the combination of omics modalities used, the number of features after preprocessing (*n_feat_samples*), the number of labeled samples (*n_labeled*), and validation metrics including Mean Squared Error (MSE), Root Mean Squared Error (RMSE), Mean Absolute Error (MAE), and the coefficient of determination ($R^2$).

## Single-Modality Performance

PLX-4720 sensitivity analysis based on individual omics layers shows substantial variability in predictive accuracy. The epigenomic modality provides the best performance among single-omics models, with a validation $R^2$ of 0.462 and an RMSE of 1.225, indicating that DNA methylation patterns strongly contribute to cell line response to PLX-4720. Genomic, proteomic, and PPI-based models achieve moderate performance with $R^2$ values ranging from 0.08 to 0.30, whereas metabolomic and transcriptomic data alone yield lower accuracy, suggesting limited inference power when relying on a single molecular layer.

### Two-Modality Integration

Integrating two modalities generally improves prediction accuracy relative to most single-omics models. The combination of genomics + transcriptomics achieves the highest performance, with a validation $R^2$ of 0.9296 and an RMSE of 0.4397, indicating that the synergy between mutational status and gene expression provides the strongest predictive signal for PLX-4720 sensitivity. The metabolomics + transcriptomics combination also performs exceptionally well with an $R^2$ of 0.9101. Other two-omics integrations involving epigenomics, proteomics, or PPI demonstrate moderate improvements, with $R^2$ values between 0.63 and 0.76, reinforcing that pairing two layers of biological regulation yields more stable predictive representations than using each layer independently.

### Three- and Four-Modality Integration

Integrating three or four omics layers yields consistent and relatively stable performance, although gains in accuracy become limited. For example, the epigenomics + metabolomics + transcriptomics configuration achieves an $R^2$ of 0.759, while epigenomics + genomics + metabolomics + PPI reaches an $R^2$ of 0.728. Overall, three- or four-modality combinations reduce prediction error compared with several single-omics models but do not surpass the accuracy of the strongest two-modality configurations—namely genomics + transcriptomics and metabolomics + transcriptomics.

### Five-Modality Integration

Integrating five omics layers produces $R^2$ values ranging from 0.75 to 0.76, indicating that multi-layer biological data provide a comprehensive representation but do not necessarily improve prediction accuracy. For instance, the combination of epigenomics + genomics + metabolomics + PPI + transcriptomics yields an $R^2$ of 0.7567, which is lower than the best-performing two-modality models. These findings suggest that adding more modalities does not always provide additional predictive benefit, particularly when information becomes redundant or insufficiently independent.

# DISCUSSION

The results demonstrate that PLX-4720 sensitivity can be accurately predicted using a multi-omics integration approach; however, improvements in accuracy are not linear with the number of combined modalities. Although stepwise integration enriches the underlying biological context, the best performance is achieved through the integration of two specific modalities—namely genomics + transcriptomics. This configuration reaches an $R^2$ above 0.92, representing the highest accuracy among all tested models.

### Biological Interpretation of Two-Modality Integration

The strong performance of the genomics + transcriptomics combination reflects the mechanism of action of PLX-4720 as a BRAF inhibitor. Response to PLX-4720 is influenced by:

- upstream mutation status in BRAF or RAS (genomics), and
- downstream pathway activation (transcriptomics).

Mutations alone cannot fully explain signaling rewiring without considering gene expression, while gene expression without mutational context may obscure driver-related effects. Thus, integrating these two regulatory layers captures MAPK pathway status more comprehensively, consistent with principles of oncogene addiction and sensitivity to BRAF inhibitors.

## Role of Other Modalities

Epigenomics emerges as the strongest single-modality predictor, indicating that DNA methylation contributes to heterogeneity in drug response. When combined with transcriptomics, proteomics, or PPI data, epigenomics offers moderate accuracy gains, reflecting its role in regulating gene expression and protein activity states.

Metabolomics and PPI data have limited predictive value on their own but enhance accuracy when integrated into two- to five-layer models. These non-transcriptomic modalities may serve as biochemical and network-level contextual enhancers within the predictive framework.

## Diminishing Returns in Higher-Level Integration

Although integrating five modalities yields the most comprehensive biological representation, predictive accuracy does not surpass that of the best two-modality combinations. This phenomenon may arise from feature redundancy, increased noise, and sample-size imbalance across modalities. When additional layers do not contribute independent signal, model complexity increases without corresponding gains in inference capability. Methodologically, these findings emphasize that selecting the right modalities is more important than maximizing the number of modalities in pharmacogenomic modeling.

## Implications for Pharmacogenomics

Models integrating two complementary omics layers—such as genomics + transcriptomics—offer substantial practical value because they:

(a) deliver maximal predictive accuracy,
(b) require fewer laboratory assays, and
(c) maintain lower model complexity than five-modality integrations.

These results imply that a selective, modality-focused strategy may be more efficient than extensive multi-omics integration for predicting drug response, prioritizing cell lines, or generating hypotheses regarding resistance mechanisms.

### Limitations and Future Directions

The sample size is relatively small (18–28 cell lines per configuration), making models susceptible to variability. This limitation is common in pharmacogenomic studies of cancer cell lines and highlights the need for external validation (Wang et al., 2021). Additionally, the dataset consists of **baseline static molecular profiles**, without post-treatment dynamics—despite the fact that resistance to BRAF inhibitors is often adaptive. Integrating dynamic or time-course omics could improve interpretability.

Although deep learning models are traditionally viewed as "black boxes," attention mechanisms and interpretability tools such as SHAP can help link predictions to key biological features (Zhang et al., 2021). For future research, several logical directions include:

- incorporating drug chemical structure, as implemented in PASO (Wu et al., 2025),
- leveraging transfer learning or multitask learning (Uyar et al., 2021),
- integrating pathway-level or network module information (Zhang et al., 2021).

Overall, this study reinforces attention-based multi-omics integration as a relevant and powerful strategy for modeling PLX-4720 sensitivity, further solidifying multidimensional omics as a foundational pillar of precision pharmacogenomics.

# CONCLUSION

This study demonstrates that PLX-4720 sensitivity in cancer cell lines can be accurately predicted using a multi-omics integration approach. Predictive performance is not determined solely by the number of modalities combined, but rather by the biologically relevant combinations of layers. Integration of two modalities—particularly genomics and transcriptomics—produces the most accurate models, achieving validation $R^2$ values above 0.92 and outperforming all three-, four-, and five-modality configurations. These findings highlight that driver mutation status and transcriptional activation of the MAPK pathway are key determinants of PLX-4720 sensitivity, consistent with the mechanism of BRAF inhibition.

Epigenomics stands out as the most informative single modality, underscoring the importance of epigenetic regulation in drug response heterogeneity. While integrating three to five omics layers improves stability and biological context, it does not always enhance predictive accuracy due to potential information redundancy, differences in data depth, or cross-modality sample imbalance. Therefore, careful modality selection is more important than simply increasing the number of omics layers.

Practically, these results have important implications for precision pharmacogenomics. Drug response prediction models do not require extensive multi-modality integration; instead, combining two complementary layers can deliver maximal performance with lower data requirements and reduced experimental costs. This study supports the use of genomics–transcriptomics integration as an efficient strategy for prioritizing cell lines, identifying oncogene dependencies, and accelerating predictive analyses for BRAF inhibitor response.